\documentclass{aastex631}
\pdfoutput=1 
\usepackage{amsmath}
\usepackage[T1]{fontenc}
\usepackage{url}
\usepackage{graphicx}
\usepackage[caption=false]{subfig}

\usepackage[caption=false]{subfig}

\begin{document}

\title{Pre-merger sky localization of gravitational waves from  binary neutron star mergers using deep learning}

\author[0000-0001-8700-3455]{Chayan Chatterjee}
\affiliation{Department of Physics, OzGrav-UWA, The University of Western Australia,\\ 35 Stirling Hwy, Crawley, Western Australia 6009, Australia}

\author[0000-0001-7987-295X]{Linqing Wen}
\affiliation{Department of Physics, OzGrav-UWA, The University of Western Australia,\\ 35 Stirling Hwy, Crawley, Western Australia 6009, Australia}





\begin{abstract}

The simultaneous observation of gravitational waves (GW) and prompt electromagnetic counterparts from the merger of two neutron stars can help reveal the properties of extreme matter and gravity during and immediately after the final plunge. Rapid sky localization of these sources is crucial to facilitate such multi-messenger observations. Since GWs from binary neutron star (BNS) mergers can spend up to 10-15 mins in the frequency bands of the detectors at design sensitivity, early warning alerts and pre-merger sky localization can beg achieved for sufficiently bright sources, as demonstrated in recent studies. In this work, we present pre-merger BNS sky localization results using \texttt{GW-SkyLocator}, a deep learning model capable of inferring sky location posterior distributions of GW sources at orders of magnitude faster speeds than standard Markov Chain Monte Carlo methods. We test our model's performance on a catalog of simulated injections from Sachdev et al. (2020), recovered at 0-60 secs before merger, and obtain comparable sky localization areas to the rapid localization tool \texttt{BAYESTAR}. These results show the feasibility of our model for pre-merger sky localization and the possibility of follow-up observations for precursor emissions from BNS mergers.

\end{abstract}




\section{Introduction} \label{sec:intro}
The first direct detection of GWs from a merging binary black hole (BBH) system was made in 2015 (\cite{GW150914}), which heralded a new era in astronomy. Since then the LIGO-Virgo-KAGRA (LVK) Collaboration (\cite{LIGO,Virgo,KAGRA}) has made more than 90 detections of GWs from merging compact binaries (\cite{GWTC3}), including two confirmed detections from merging binary neutron stars (BNS) and two from mergers of neutron star-black hole (NSBH) binaries (\cite{GWTC3,Abbott_2021}). The first detection of GWs from a BNS merger on August 17th, 2017 (GW170817) along with its associated electromagnetic (EM) counterpart revolutionized the field of multi-messenger astronomy (\cite{GW170817}). This event involved the joint detection of the GW signal by LIGO and Virgo, and the prompt short gamma-ray burst (sGRB) observation by the Fermi-GBM and INTEGRAL space telescopes (\cite{GW170817_1, GW170817_2}) $\sim $ 2 secs after the merger. This joint observation of GWs and sGRB, along with the observations of EM emissions at all wavelengths for months after the event had a tremendous impact on astronomy, leading to -- an independent measurement of the Hubble Constant (\cite{HubbleConstant}), new constraints on the neutron star equation of state (\cite{EOS}) and confirmation of the speculated connection between sGRB and kilonovae with BNS mergers (\cite{GW170817_1}).

While more multi-messenger observations involving  GWs are certainly desirable, the typical delays between a GW detection and the associated GCN alerts, which is of the order of a few minutes (\cite{Magee1}), makes such joint discoveries extremely challenging. Therefore, an advance warning system with pre-merger sky localization capabilities will open up new modes of observation to search for prompt emission in wavelengths outside of gamma rays, like ultraviolet, optical and near-infrared (\cite{multimessenger1,multimessenger2, GOTO}) from the early kilonova using facilities like Swift/XRT and LOFAR.

In recent years, several studies have shown that for a fraction of BNS events, it will be possible to issue alerts up to 60 secs before merger (\cite{Magee1, Sachdev, Kovalam, Nitz2020}). Such early-warning detections, along with pre-merger sky localizations will facilitate rapid EM follow-up of prompt emissions (\cite{Pre-mergerEM1, Pre-mergerEM2}). The observations of optical and ultraviolet emissions prior to mergers are necessary for understanding r-process nucleosynthesis (\cite{r-process}) and the physics governing emissions of shock-heated ejecta (\cite{shock-heated_ejecta}). Prompt X-ray emission can reveal the final state of the remnant (\cite{X-ray1, X-ray2, X-ray3}) post merger, and early radio observations can reveal the nature of pre-merger magnetosphere interactions (\cite{radio}), and help test theories connecting BNS mergers with fast radio bursts (\cite{BNS-FRB1, BNS-FRB2, BNS-FRB3}). 

In the last three LVK observation runs, five GW low-latency detection pipelines have processed data and sent out alerts in real-time. These pipelines are GstLAL (\cite{GSTLAL}), SPIIR (\cite{SPIIR}), PyCBC (\cite{PyCBCpipeline}), MBTA (\cite{MBTA}), and cWB (\cite{cWB}). These pipelines also use the rapid localization tool, \texttt{BAYESTAR} (\cite{Bayestar}) to produce prompt estimates of the sky directions of GW candidate sources in around one second.

Of the five search pipelines, the first four use the technique of matched filtering (\cite{ShaunHooper}) to identify real GW signals in detector data, while cWB uses a coherent analysis to search for burst signals in detector data streams. In 2020, an end-to-end mock data challenge (\cite{Magee1}) was conducted by the GstLAL and SPIIR search pipelines and successfully demonstrated their feasibility to send pre-merger alerts (\cite{Magee1}). This study also estimated the expected rate of BNS mergers and their sky localization areas using \texttt{BAYESTAR} from a four detector network consisting of LIGO Hanford (H1), LIGO Livingston (L1), Virgo (V1) and KAGRA in O4 detector sensitivity. In a previous study, Sachdev et al. (2020) (\cite{Sachdev}) showed early warning performance of the GstLAL pipeline over a month of simulated data with injections. Their study suggested that alerts could be issued 10s (60 s) before merger for 24 (3) BNS systems over the course of one year of observations of a three-detector Advanced network operating at design sensitivity. These findings were in broad agreement with the estimates of Cannon et al. (2012) (\cite{Cannon2012}) on the rates of early warning detections at design sensitivity. Sky localization results for these studies, obtained at various number of seconds before merger using \texttt{BAYESTAR} indicated that around one event will be simultaneously detected pre-merger and localized within 100 deg$^{2}$ in the sky, based on current BNS merger rate estimates. 


During the third observation run of LVK, \texttt{BAYESTAR}'s run time averaged to about 1 second over all events, which is several orders of magnitude faster than full parameter estimation runs that typically took between 6 hours and 6 days. Similar localization speeds to \texttt{BAYESTAR} can be obtained using machine learning techniques, as demonstrated in Chatterjee et al. (2022), (\cite{GW-SkyLocator}) which motivates future efforts towards building similar deep learning-based techniques for full parameter estimation of all compact binary sources with a latency of seconds or less.

In this paper, we report pre-merger sky localization using deep learning for the first time. We obtain our results using \texttt{GW-SkyLocator}, a normalizing flow model (\cite{NormalizingFlow, MAF1, MAF2}) for sky localization of all types of compact binary coalescence sources (\cite{GW-SkyLocator}). We test our model on simulated BNS events from the injection catalog in Sachdev et al. (2020) (\cite{Sachdev}), that consists of signals detected at 0 to 60 secs before merger using the GstLAL search pipeline. We compare our sky localization performance with \texttt{BAYESTAR} and find that our localization contours have comparable sky contour areas with \texttt{BAYESTAR}, at an inference speed of just a few milliseconds using a NVIDIA P100 12 GB PCIe GPU. 

The paper is divided as follows: we briefly describe our normalizing flow model in Section 2. In Section 3, we describe the details of the simulations used to generate the training and test sets. In Section 4, we desribe our architecture of \texttt{GW-SkyLocator}. In Section 5, we discuss results obtained using our network on the dataset from Sachdev et al. (2020) (\cite{Sachdev}). Finally, we discuss future directions of this research in Section 6.

\section{Method} \label{sec:Method}
Our neural network, \texttt{GW-SkyLocator} is based on a class of deep neural density estimators called normalizing flow, the details of which is provided in (\cite{GW-SkyLocator}). \texttt{GW-SkyLocator} consists of three main components: (i) the normalizing flow, specifically, a Masked Autoregressive Flow (MAF) (\cite{MAF1, MAF2}) network, (ii) a ResNet-34 model (\cite{ResNet-34}) that extracts features from the complex signal-to-noise (SNR) time series data which is obtained by matched filtering GW strains with BNS template waveforms, and (iii) a fully connected neural network whose inputs are the intrinsic parameters (component masses and z-component of spins) of the templates used to generate the SNR time series by matched filtering. The architecture of our model is shown in Figure 1. The features extracted by the ResNet-34 and fully connected networks from the SNR time series ($\rho(t)$) and best-matched intrinsic parameters ($ {\theta}_{in}$) respectively, are combined into a single feature vector and passed as a conditional input to the MAF. The MAF is a normalizing flow with a specific architecture, that transforms a simple base distribution (a multi-variate Gaussian) $ z \sim p(z)$ into a more complex target distribution $x \sim p(x) $ which in our case, is the posterior distribution of the right ascension ($\alpha$) and declination angles ($\delta$) of the GW events, given the SNR time series and intrinsic parameters $p(\alpha, \delta|\rho(t),  {\theta}_{in})$. 

This mapping is learnt by the flow during training using 
the method of maximum likelihood, and can be expressed as:

\begin{equation}
    p(x) = \pi(z)\left|\text{det}\frac{\partial f(z)}{\partial z}\right|^{-1},
\end{equation}

If $z $ is a random sample drawn from the base distribution $\pi(z) $, and $f $ is the invertible transformation parametrized by the normalizing flow, then the new random variable obtained after the transformation is $x = f(z)$. The transformation, $f $ can be made more flexible and expressive by stacking a chain of transformations together as follows:

\begin{equation}
    x_{k}=f_{k}\circ \ldots \circ f_{1}\left( z_{0}\right) 
\end{equation}

This helps the normalizing flow learn arbitrarily complex distributions, provided each of the transformations are invertible and the Jacobians are easy to evaluate. Neural posterior estimation (NPE) (\cite{NPE1, NPE2, NPE3}) techniques, including normalizing flows and conditional variational autoencoders have been used to estimate posterior distribution of BBH source parameters with high accuracy and speed (\cite{Green, Gabbard, Chua}). Chatterjee et al. (2022) (\cite{GW-SkyLocator}) used a normalizing flow to demonstrate rapid inference of sky location posteriors for all CBC sources for the first time. This work shows the first application of deep learning for pre-merger BNS sky localization and is an extension of the model introduced in \cite{GW-SkyLocator} 

\begin{figure*}
  \centering
  \includegraphics[scale=0.60]{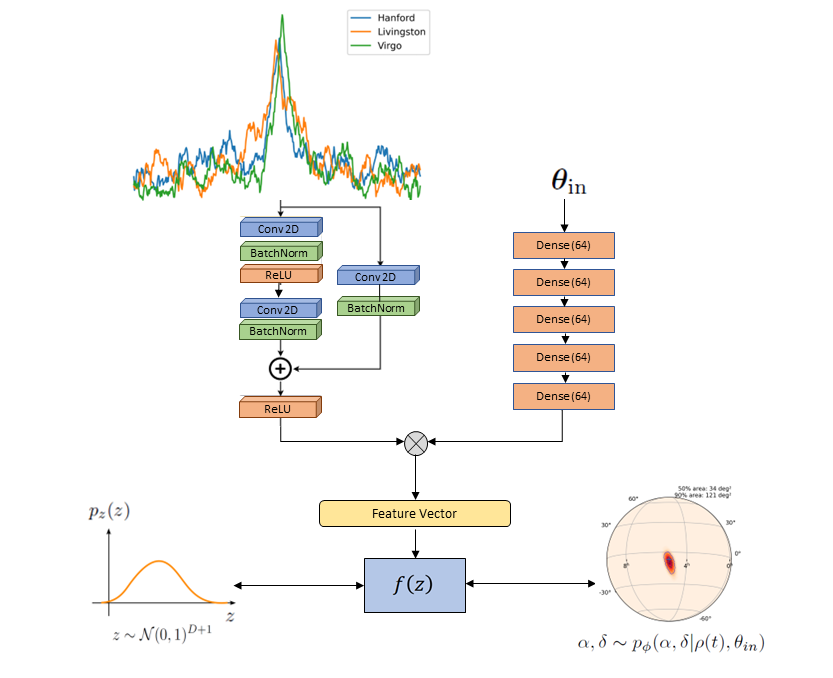}
\caption{\label{fig:Model_architecture} Architecture of our model, \texttt{GW-SkyLocator}. The input data, consisting of the SNR time series, $\rho(t) $ and intrinsic parameters, $ {\theta}_{in}$ are provided to the network through two separate channels: the ResNet-34 channel (only one ResNet block is shown here) and the multi-layered fully connected (Dense) network respectively. The features extracted by $\rho(t) $ and $ {\theta}_{in}$ are then combined and provided as conditional input to the main component of \texttt{GW-SkyLocator} - the Masked Autoregressive Flow (MAF) network , denoted by $f(z)$. The MAF draws samples, $z $, from a multivariate Gaussian, and learns a mapping between $z $ to ($\alpha $, $\delta $), which are the right ascension and declination angles of the GW events.}

%

\end{figure*}

\section{\label{sec:level1} Data Generation}

We train six different versions of \texttt{GW-SkyLocator} with distinct training sets ($\rho^{i}(t)$, $ {\theta}_{in}^{i}$) for each ``negative latency", ${i = 0, 10, 14, 28, 44, 58}$ secs before merger. Our training and test set injections parameters were sampled from the publicly available injection dataset used in Sachdev et al. (2020) \cite{Sachdev}. These $ {\theta}_{in}^{i}$ parameters were used to first simulate the BNS waveforms using the \texttt{SpinTaylorT4} approximant (\cite{SpinTaylorT4}) and then injected into Gaussian noise with advanced LIGO power spectral density (PSD) at design sensitivity (\cite{PSD}) to obtain the desired strains. The SNR time series, $\rho^{i}(t)$, was then obtained by matched filtering the simulated BNS strains with template waveforms. 

For generating the training sets, the template waveforms for matched filtering were simulated using the \textit{optimal} parameters, which have the exact same values as the injection parameters used to generate the detector strains. The SNR time series obtained by matched filtering the strains with the optimal templates, $\rho^{i}_{\text{opt}}(t)$, and the optimal intrinsic parameters, $ {\theta}_{in}^{i, \text{opt}}$, were then  used as input to our network during the training process.
For testing, the template parameters were sampled from publicly available data by Sachdev et al. (2020) (\cite{Sachdev}). These parameters correspond to the parameters of the maximum likelihood or `best-matched' signal template recovered by the GstLAL matched-filtering search pipeline. Therefore the values of the intrinsic parameters $\hat{\theta}_{in}^{i}$ used during testing are close to, but is not the exact same as $ {\theta}_{in}^{i, \text{opt}}$. Similarly, the SNR time series $\rho^{i}(t)$ is not exactly similar to the optimal $\rho^{i}_{\text{opt}}(t)$, and has a slightly lower peak amplitude than the corresponding $\rho^{i}_{\text{opt}}(t)$ peak because of the small mismatch between the injection parameters and the best-matched template waveform parameters. 

While our injections have the same parameter distribution as (\cite{Sachdev}), we only choose samples with network SNRs lying between 8 and 40 at each negative latency for this study. 
This is because in the dataset from (\cite{Sachdev}), injection samples with SNR $>$ 40 are much fewer in number compared to samples between SNR 8 and 40, which means for models trained on data with parameters from (\cite{Sachdev}), there exists very few training examples with SNR $>$ 40 to learn from. One can potentially generate training sets with uniform SNR distribution spanning the entire range in (\cite{Sachdev}) which corresponds to a uniform distribution of sources in comoving volume up to a redshift of 0.2, this would be require generating an infeasibly large number of training samples for each negative latency. Also, events detected with SNR $>$ 40 are expected to be exceptionally rare, even at design sensitivities of advanced LIGO and Virgo, which is why we choose to ignore them for this study. We therefore generate samples with uniformly distributed SNRs between 8 and 40 for training, while our test samples have the same SNR distribution as (\cite{Sachdev}) between 8 and 40.

\section{Network Architecture} \label{sec:Network Architecture}

In this section, we describe the architecture of the different components of our model. The MAF is implemented using a neural network that is designed to efficiently model conditional probability densities. This network is called Masked Autoencoder for Density Estimation (MADE) (\cite{MADE}). We stack 10 MADE blocks together to make a sufficiently expressive model, with each MADE block consisting of 5 layers with 256 neurons in each layer. In between each pair of MADE networks, we use batch normalization to stabilize training. We use a ResNet-34 model (\cite{ResNet-34}), that is constructed using 2D convolutional and MaxPooling layers with skip connections, (\cite{ResNet-34}) to extract features from the SNR time series data. The real and imaginary parts of the SNR time series are stacked vertically to generate a two dimensional input data stream for each training and test sample. The initial number of kernels for the convolutional layers of the ResNet model is chosen to be 32, which is doubled progressively through the network (\cite{ResNet-34}). The final vector of features obtained by the ResNet are combined with the features extracted from the intrinsic parameters, $ {\theta}_{in}^{i}$, by the fully-connected network, consisting of 5 hidden layers with 64 neurons in each hidden layer. The combined feature vector is then passed as a conditional input to the MAF which learns the mapping between the base and target distributions during training. 

\section{Results}
In this section, we describe our test results at each negative latency. To obtain our results, we used HEALPix (\cite{HEALPix}) to divide the sky into equal area pixels over which we evaluated the probability density of the sky location of our injection samples. We then computed the areas of the 90\% credible intervals and searched probability contours from our model's predictions and compared with \texttt{BAYESTAR}. The searched probability contour corresponds to the smallest region in the sky that needs to be searched to locate the true sky direction of the event. In order to obtain the searched probability contour of an event, we rank the sky pixels by descending probability density and then sum up the probabilities of the pixels until we reach the pixel that contains the injected sky location. The contour encompassing these pixels is the searched probability and the sum of the areas of the pixels within the contour gives us the searched area. In order to evaluate the probability density, we sample 5000 $\alpha $ and $\delta $ posterior samples for each event and then train a 2D Gaussian kernel density estimator (KDE) over the samples with a bandwidth  of 0.03. The trained KDE is then used to evaluate the probability density over the sky pixels using the adaptive sampling scheme employed by \texttt{BAYESTAR} (\cite{Bayestar}). In this scheme, the posterior probability is first evaluated over $N_{side,0}$ = 16 HEALPix grids (\cite{HEALPix}), corresponding to a single sky grid area of 13.4 deg$^{2}$. The highest probability grids are then adaptively subdivided into smaller grids over which the posterior is evaluated again. This process is repeated seven times, with the highest possible resolution at the end of the iteration being $N_{side}$ = 2$^{11}$, with an area of $\sim $ 10$^{-3}$ deg$^{2}$ for the smallest grid (\cite{Bayestar}). 

Figure 2 (a) to (f) shows the histograms of the areas of the 90\% credible intervals of the predicted posterior distributions from \texttt{GW-SkyLocator} (blue) and \texttt{BAYESTAR} (orange), evaluated on the injections in (\cite{Sachdev}) with network SNRs between 8 and 40. We observe that for most of the test sets, our model predicts smaller median 90\% credible interval areas than \texttt{BAYESTAR}. Also, \texttt{BAYESTAR} shows broader tails at $<$ 100 deg$^2$, compared to \texttt{GW-SkyLocator}, especially for 0 secs, 10 secs and 15 secs before merger (Figures 2 (a), (b) and (c)). These injections, with 90\% areas $<$ 100 deg$^2$ typically have SNR $>$ 30, which show that although \texttt{GW-SkyLocator} produces smaller median 90 \% credible interval areas, it fails to match \texttt{BAYESTAR}'s precision for SNR > 30 cases. The number of injections localized with a 90\% credible interval area between 1 - 10 deg$^{2}$ by \texttt{GW-SkyLocator} is also much lower than \texttt{BAYESTAR}, although this effect is much less prominent for the other test sets.

Figures 3 (a)-(f) show the distributions of the searched areas of the injections from 0 to 58 secs before merger. 
The shape of the searched area distributions, including the median values from \texttt{GW-SkyLocator} and \texttt{BAYESTAR} are much more similar, compared to the 90\% areas. These results show that while \texttt{GW-SkyLocator} achieves similar accuracy, on average, in its estimate of the sky location, it struggles to match the same precision, described by the area of a given credible region, as the ones obtained by \texttt{BAYESTAR}. Further investigation into our training method, including use of more training data and better hyperparameter optimization methods may help improve our model's overall performance.

Figures 4 (a) and (b) show box and whisker plots for 90\% credible interval areas and searched areas obtained by \texttt{GW-SkyLocator} (blue) and \texttt{BAYESTAR} (pink) respectively. We observe that our median 90\% areas (white horizontal lines) for most of the cases are smaller than \texttt{BAYESTAR}'s.

Figures 5 (a) - (f) show Probability-Probability (P-P) plots for the injections at 0 secs, 10 secs, 15 secs, 28 secs, 44 secs and 58 secs before merger respectively. To obtain the P-P plots, we make a cumulative histogram of the credible regions corresponding to the searched probabilities described earlier, and obtain a diagonal curve for each test case. For accurate inference, the posterior densities should be self-consistent from a frequentist definition, meaning 90\% of the events should have their true locations within the 90\% credible region and so on, i.e., the P-P curve should be diagonal. The figures indicate that except for 0 secs, all the diagonal curves show some deviation from the 95\% confidence intervals (gray band) of the true diagonals (dotted lines), which could be a consequence of our choice of fixed KDE bandwidth for the posterior density evaluation. Further investigation into methods that can produce the posterior densities directly would enable us to bypass the KDE step altogether, and is expected to result in diagonal P-P curves for all these cases.


Because of the low dimensionality of our input data, training our network takes less than an hour on a NVIDIA Tesla P100 GPU. During inference, the total time taken for drawing 5000 samples, training the KDE, evaluating it over $N_{side,0}$ = 16 HEALPix grids and performing the eight rounds of adaptive refinement takes about 1.2 secs on average. Sample generation and matched filtering was implemented with a modified version of the code developed by (\cite{Gebherd}) that uses \texttt{PyCBC} software (\cite{PyCBC}). \texttt{GW-SkyLocator} was written in TensorFlow 2.4 (\cite{TensorFlow}) using the Python language.

\begin{figure}
\gridline{\fig{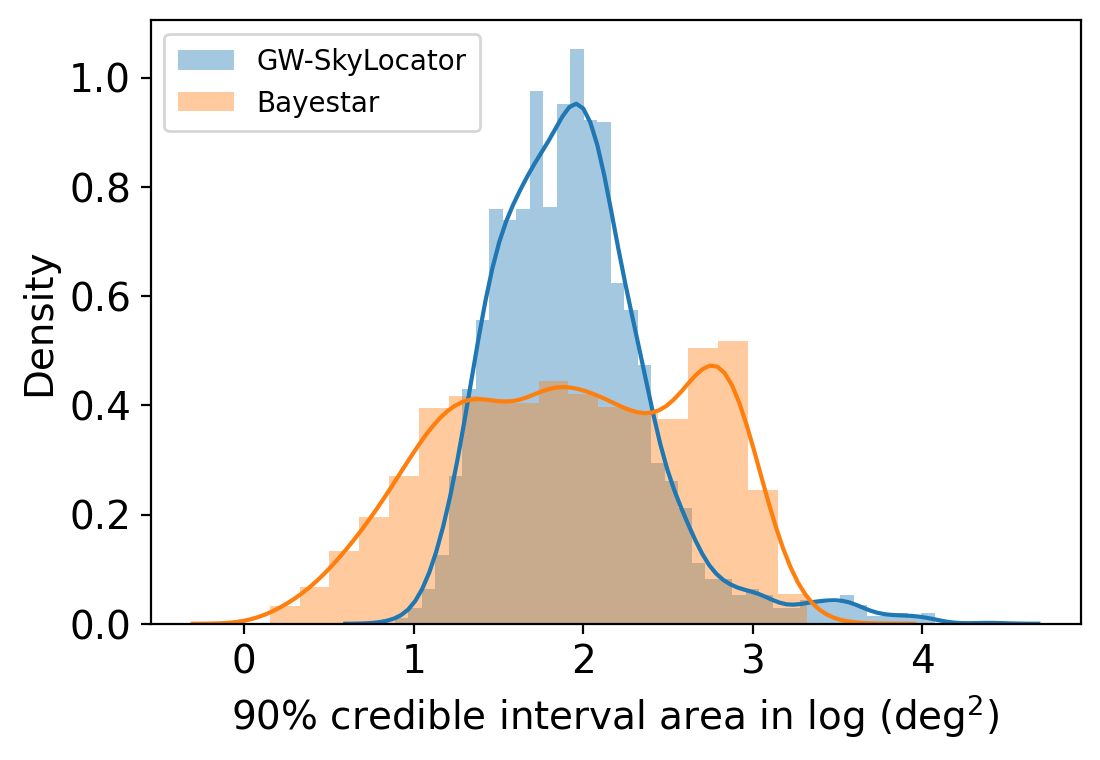}{0.32\textwidth}{(a)}
          \fig{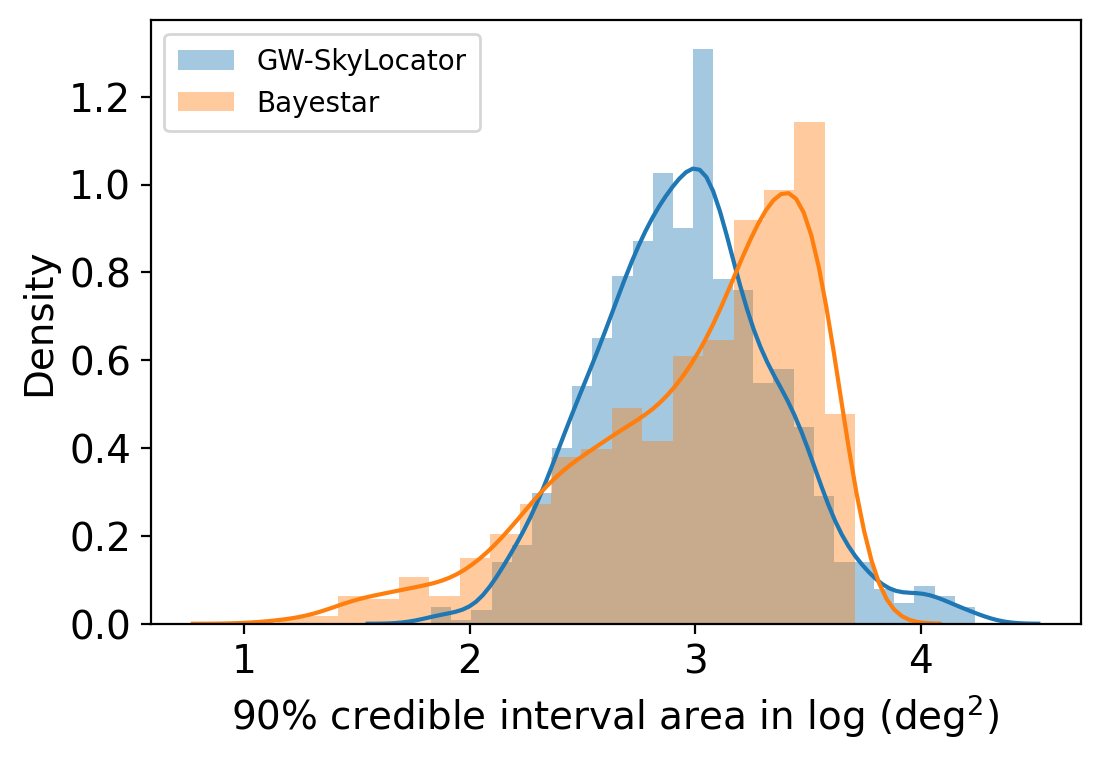}{0.32\textwidth}{(b)}
          \fig{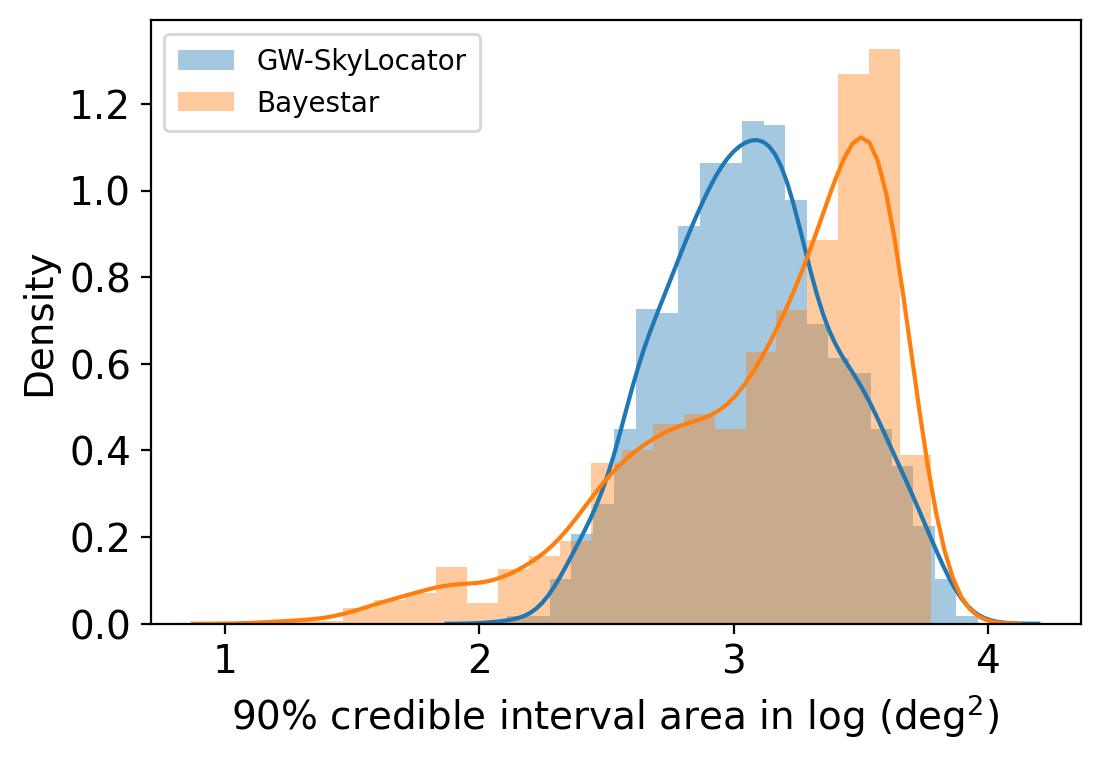}{0.32\textwidth}{(c)}}
\gridline{\fig{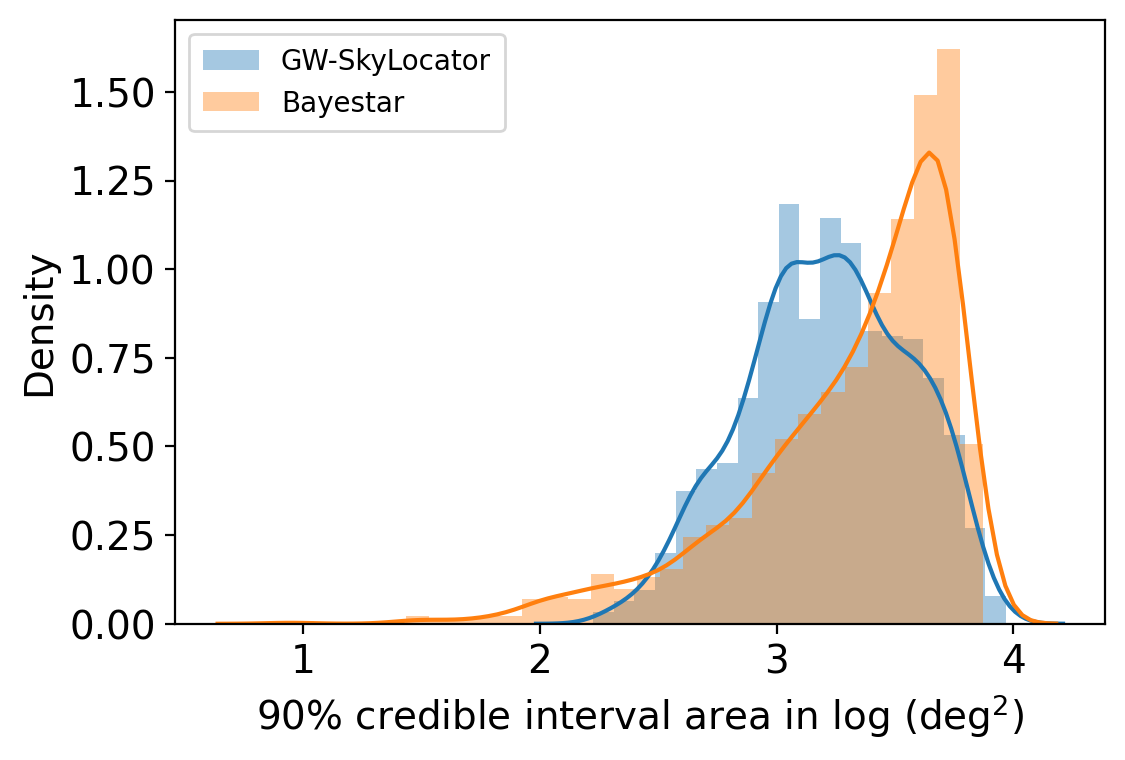}{0.32\textwidth}{(d)}
          \fig{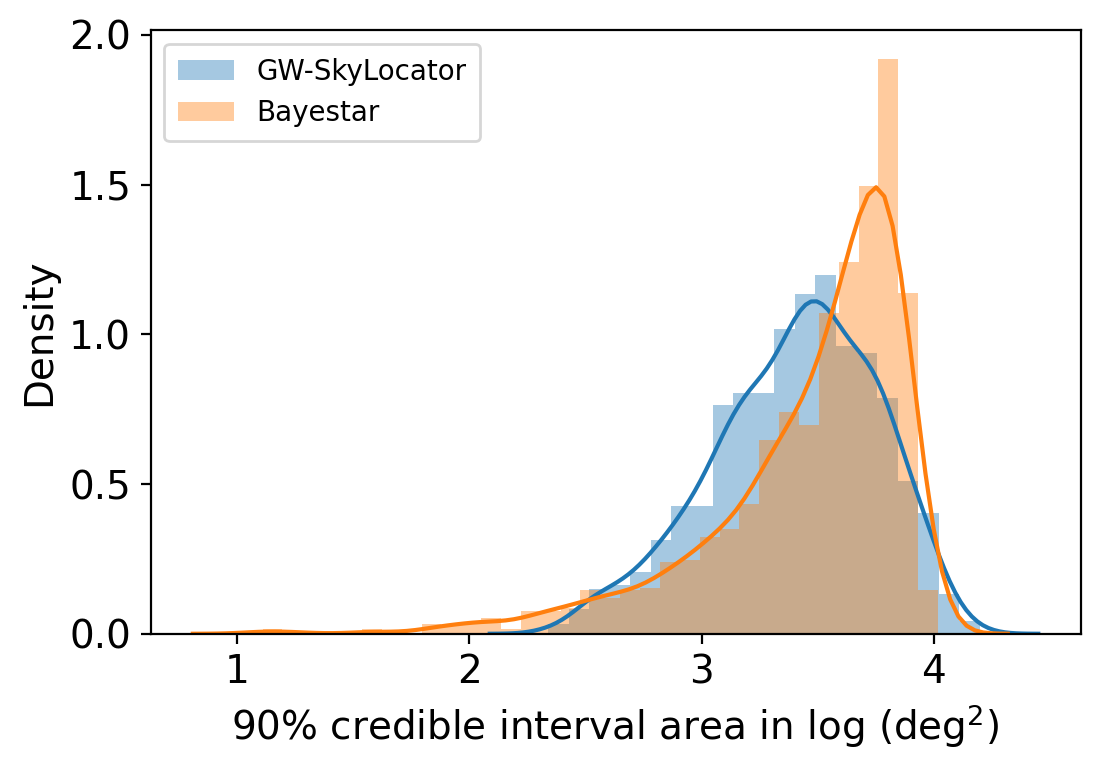}{0.32\textwidth}{(e)}
          \fig{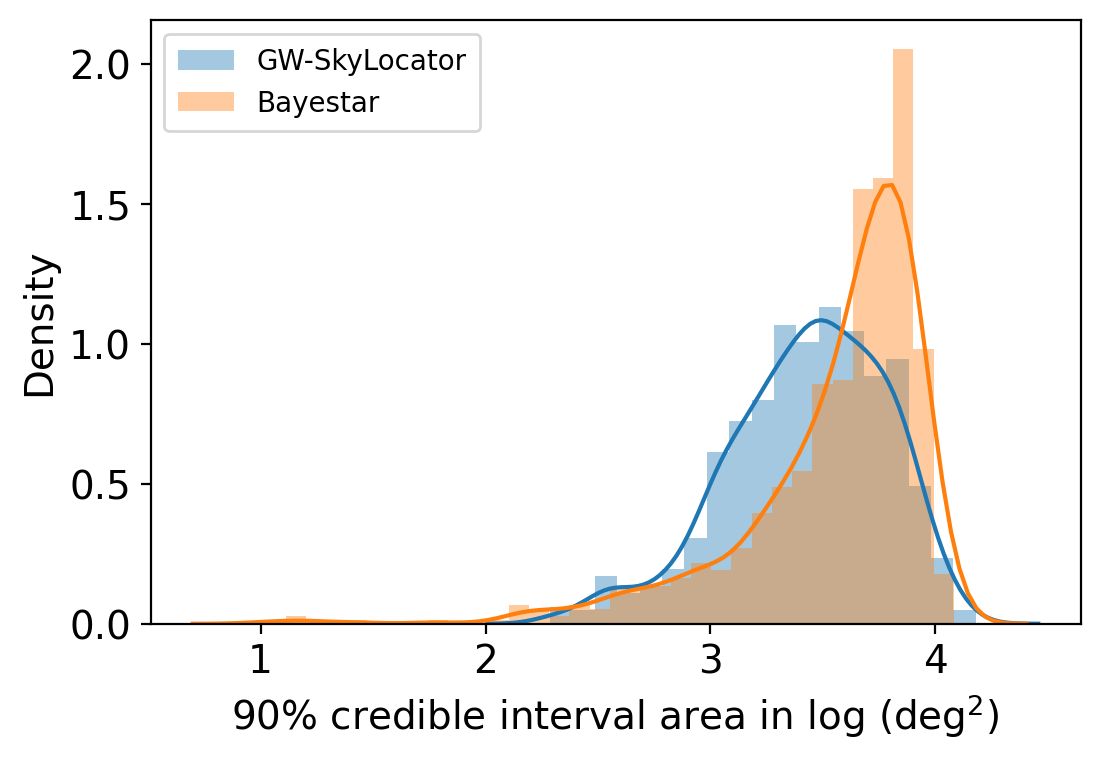}{0.32\textwidth}{(f)}}
\caption{\label{fig:90_percent} Top panel from (a) to (c): Histograms of the areas of the 90\% credible intervals of 2000 BNS events obtained from \texttt{GW-SkyLocator} (blue) and \texttt{BAYESTAR} (orange) for 0 secs, 10 secs, 15 secs before merger are shown. Bottom panel from (d) to (f): Similar histograms for 28 secs, 44 secs and 58 secs before merger are shown.}
\end{figure}

\begin{figure}
\gridline{\fig{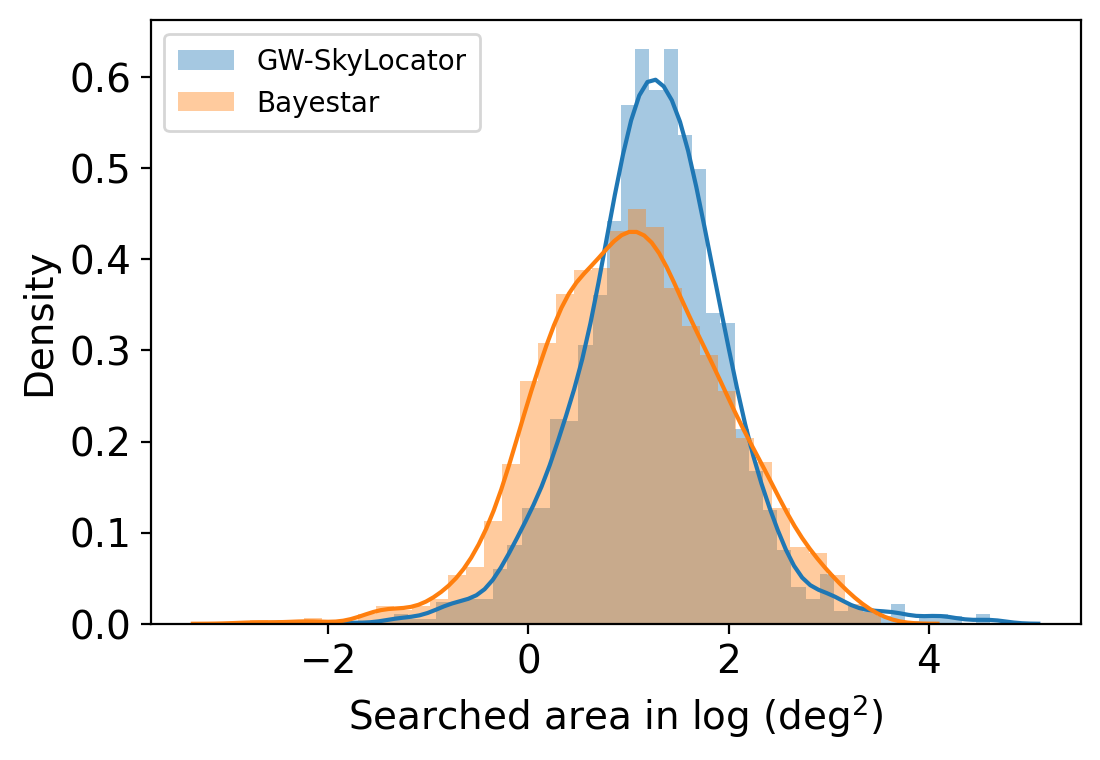}{0.32\textwidth}{(a)}
          \fig{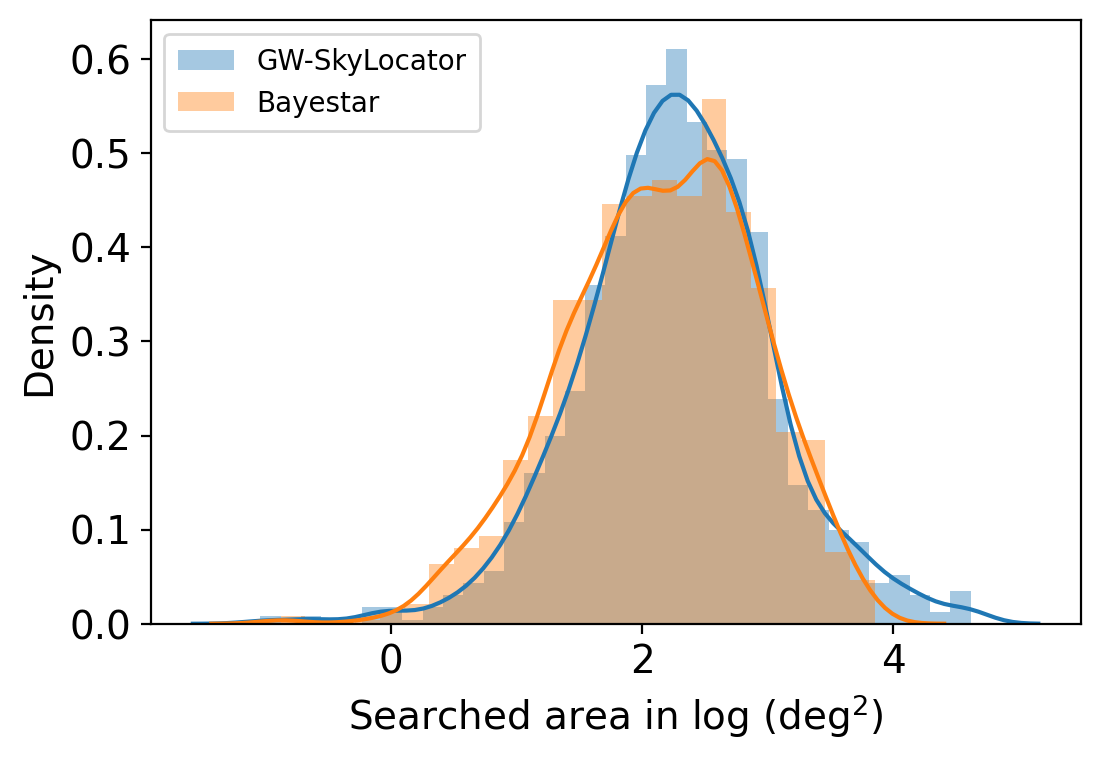}{0.32\textwidth}{(b)}
          \fig{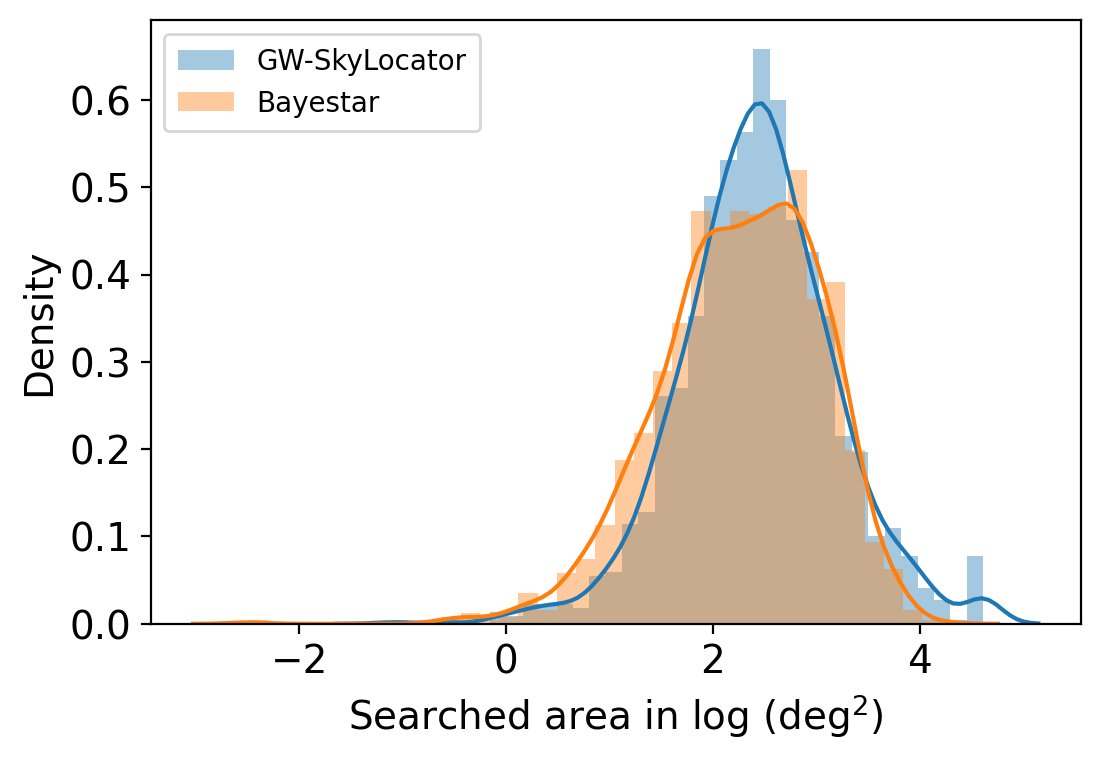}{0.32\textwidth}{(c)}}
\gridline{\fig{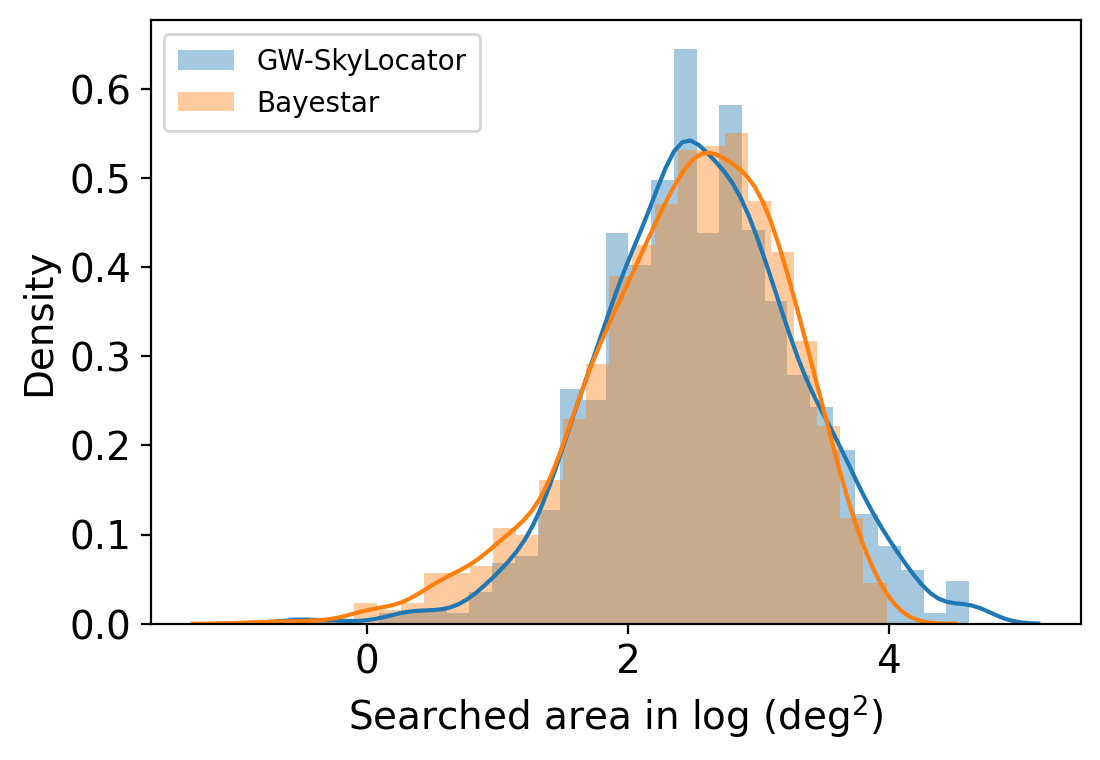}{0.32\textwidth}{(d)}
          \fig{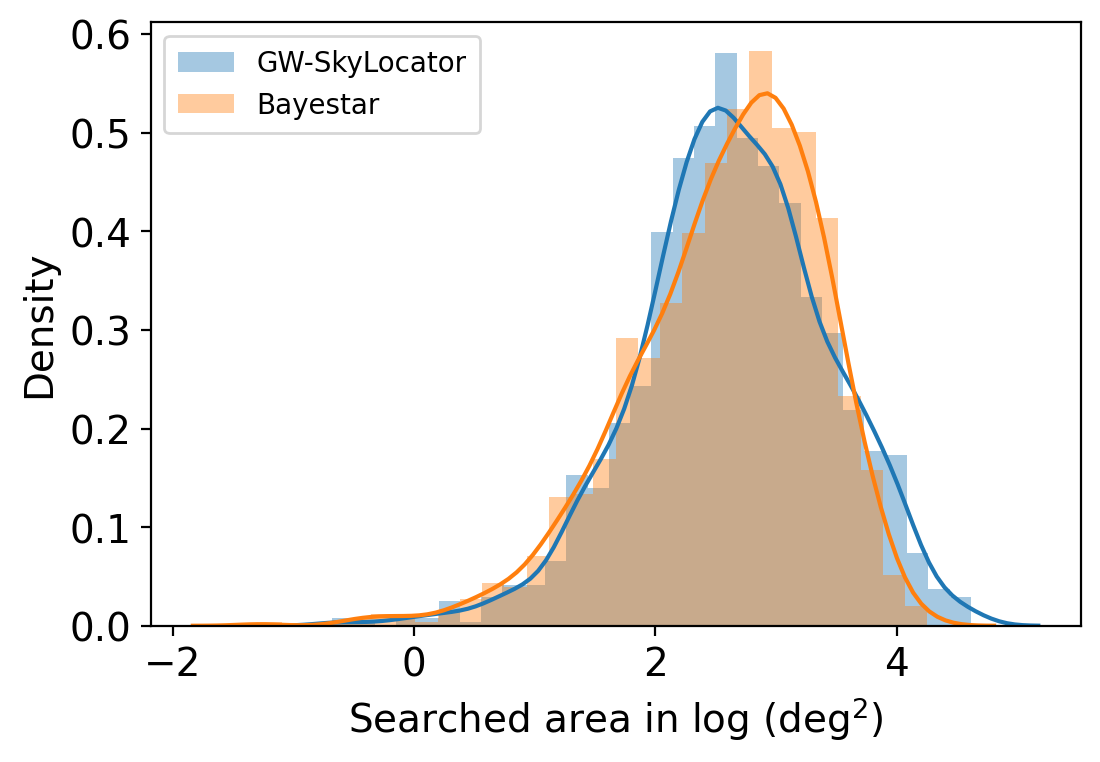}{0.32\textwidth}{(e)}
          \fig{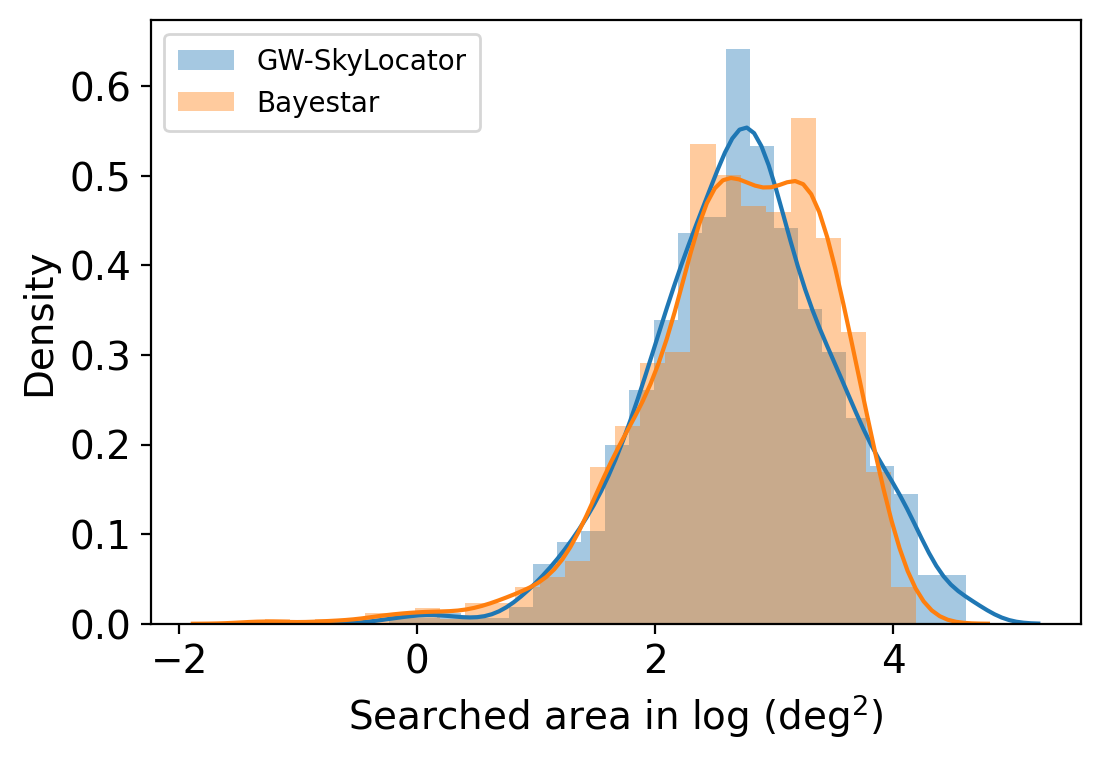}{0.32\textwidth}{(f)}}
\caption{\label{fig:Searched_area} Top panel from (a) to (c): Histograms of the searched areas of 2000 BNS events obtained from \texttt{GW-SkyLocator} (blue) and \texttt{BAYESTAR} (orange) for 0 secs, 10 secs, 15 secs before merger are shown. Bottom panel from (d) to (f): Similar histograms for 28 secs, 44 secs and 58 secs before merger are shown.}
\end{figure}

\begin{figure}
\gridline{\fig{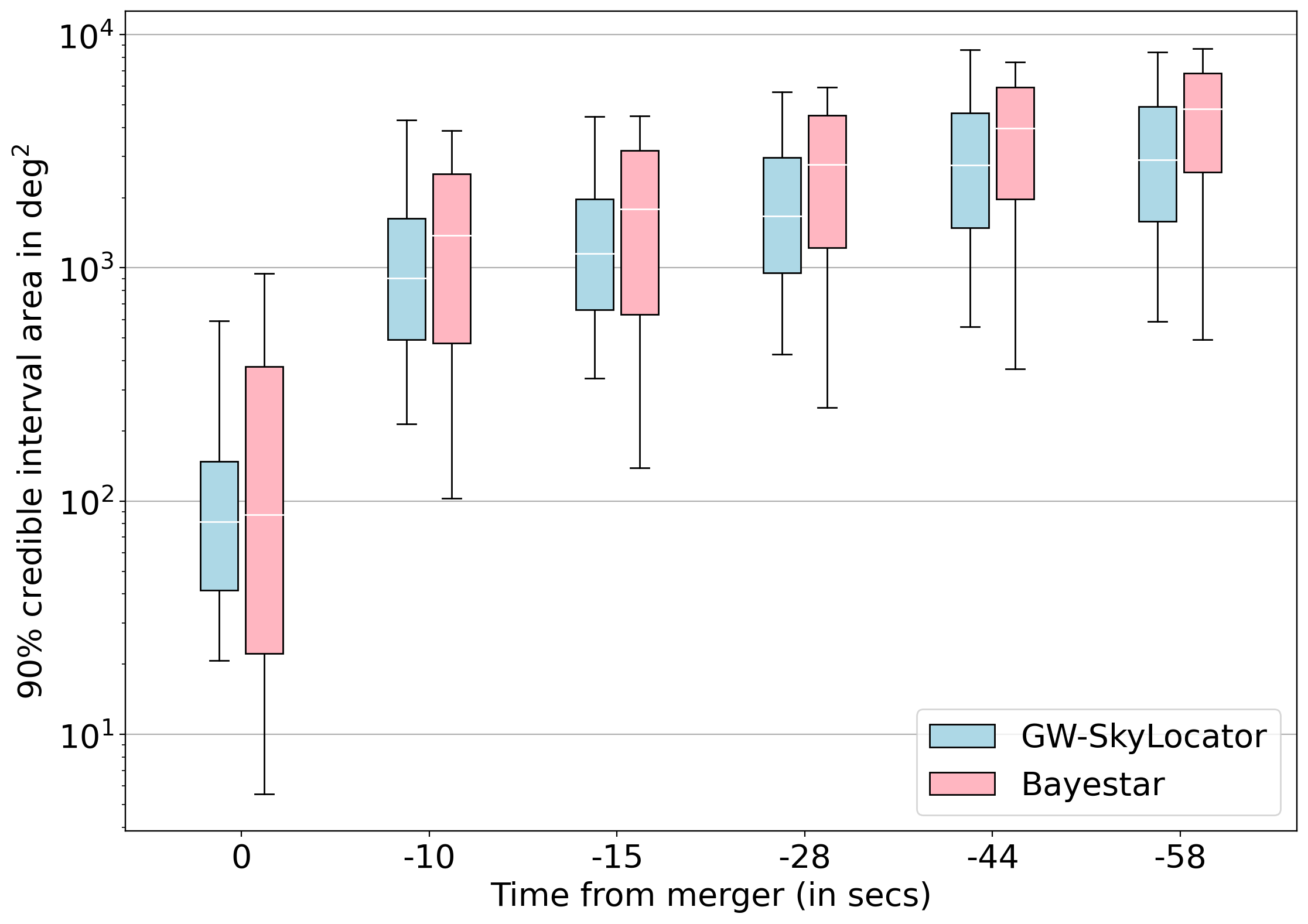}{0.45\textwidth}{(a)}
          \fig{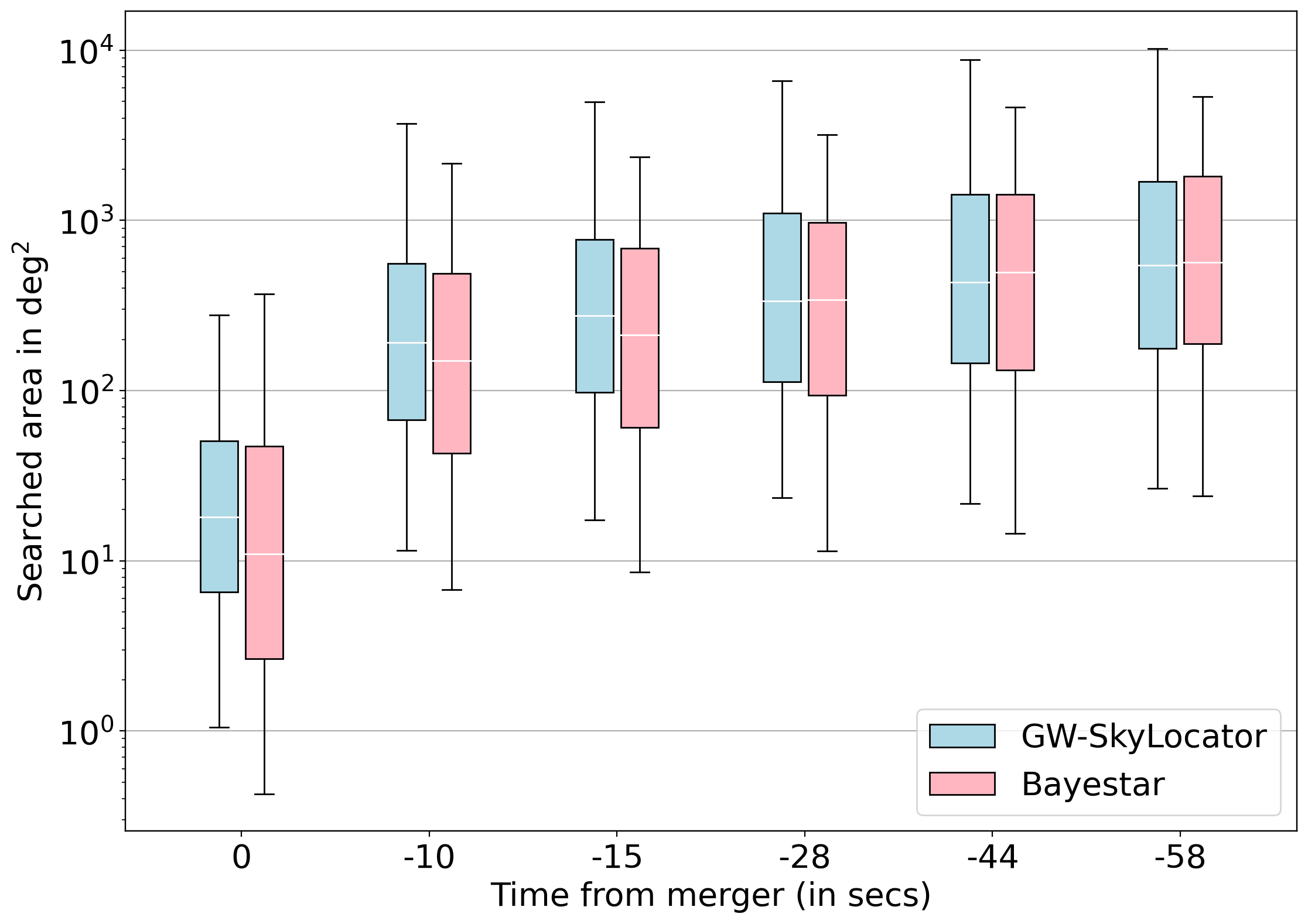}{0.45\textwidth}{(b)}}
\caption{\label{fig:Box_plots} (a) Box and whiskers plots showing the areas of the 90\% credible intervals from \texttt{GW-SkyLocator} (blue) and \texttt{BAYESTAR} (pink) at 0 secs, 10 secs, 15 secs, 28 secs, 44 secs and 58 secs before merger. The boxes encompass 95\% of the events and the whiskers extend up to the rest. The white lines within the boxes represent the median values of the respective data sets. (b) Similar box and whiskers plot as (a) for comparing searched areas from \texttt{GW-SkyLocator} (blue) and \texttt{BAYESTAR} (pink) at 0 secs, 10 secs, 15 secs, 28 secs, 44 secs and 58 secs before merger.}
\end{figure}

\begin{figure}
\gridline{\fig{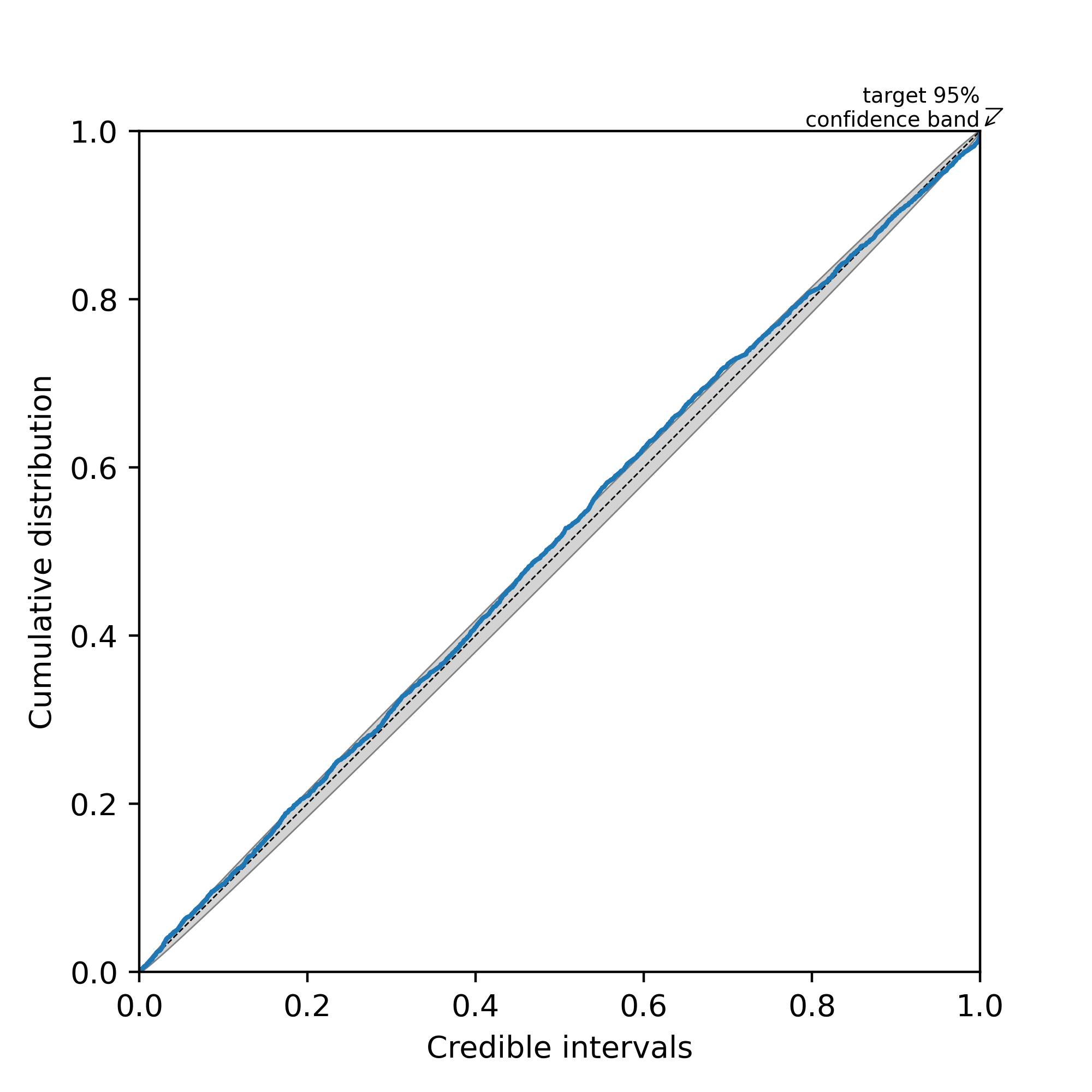}{0.32\textwidth}{(a)}
          \fig{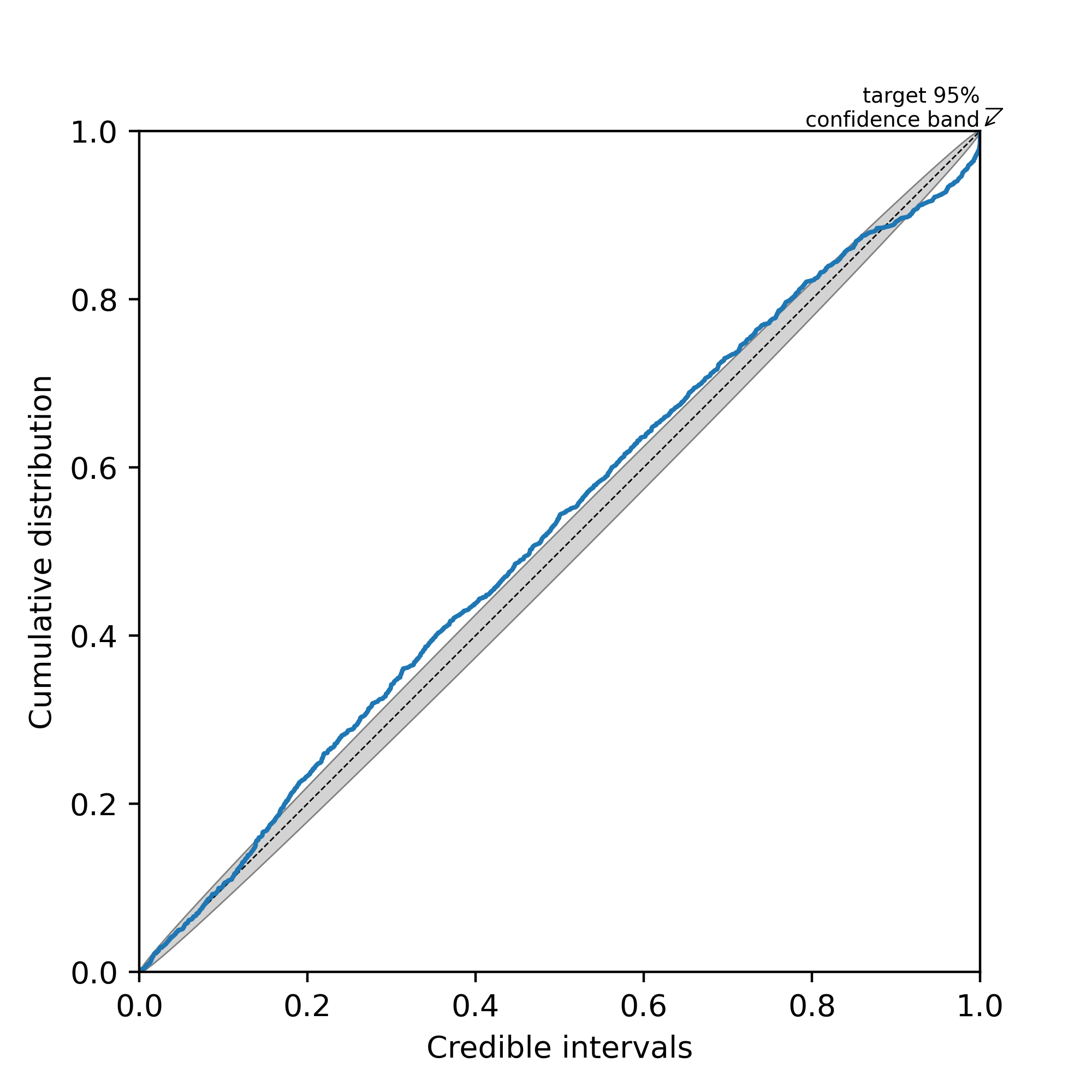}{0.32\textwidth}{(b)}
          \fig{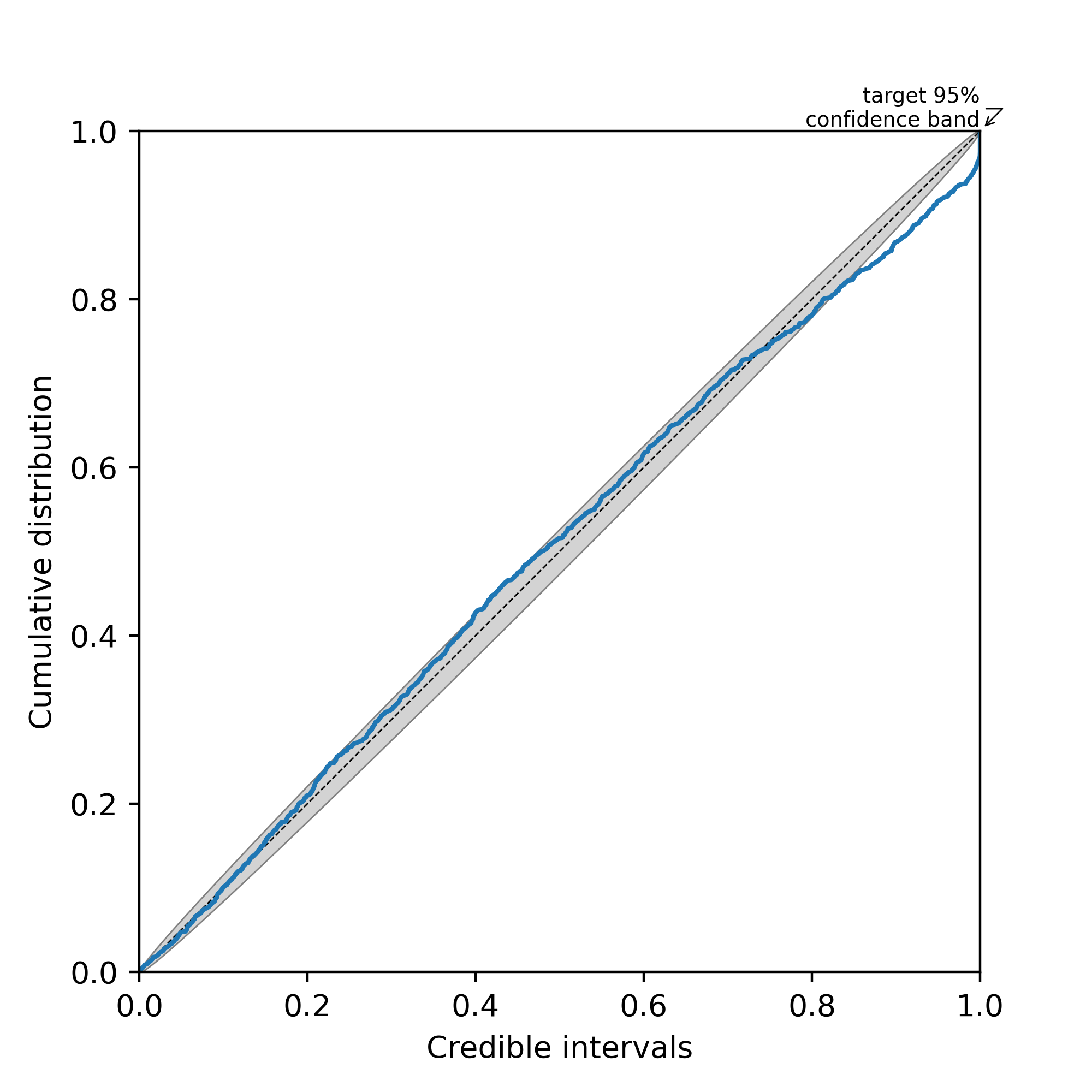}{0.32\textwidth}{(c)}}
\gridline{\fig{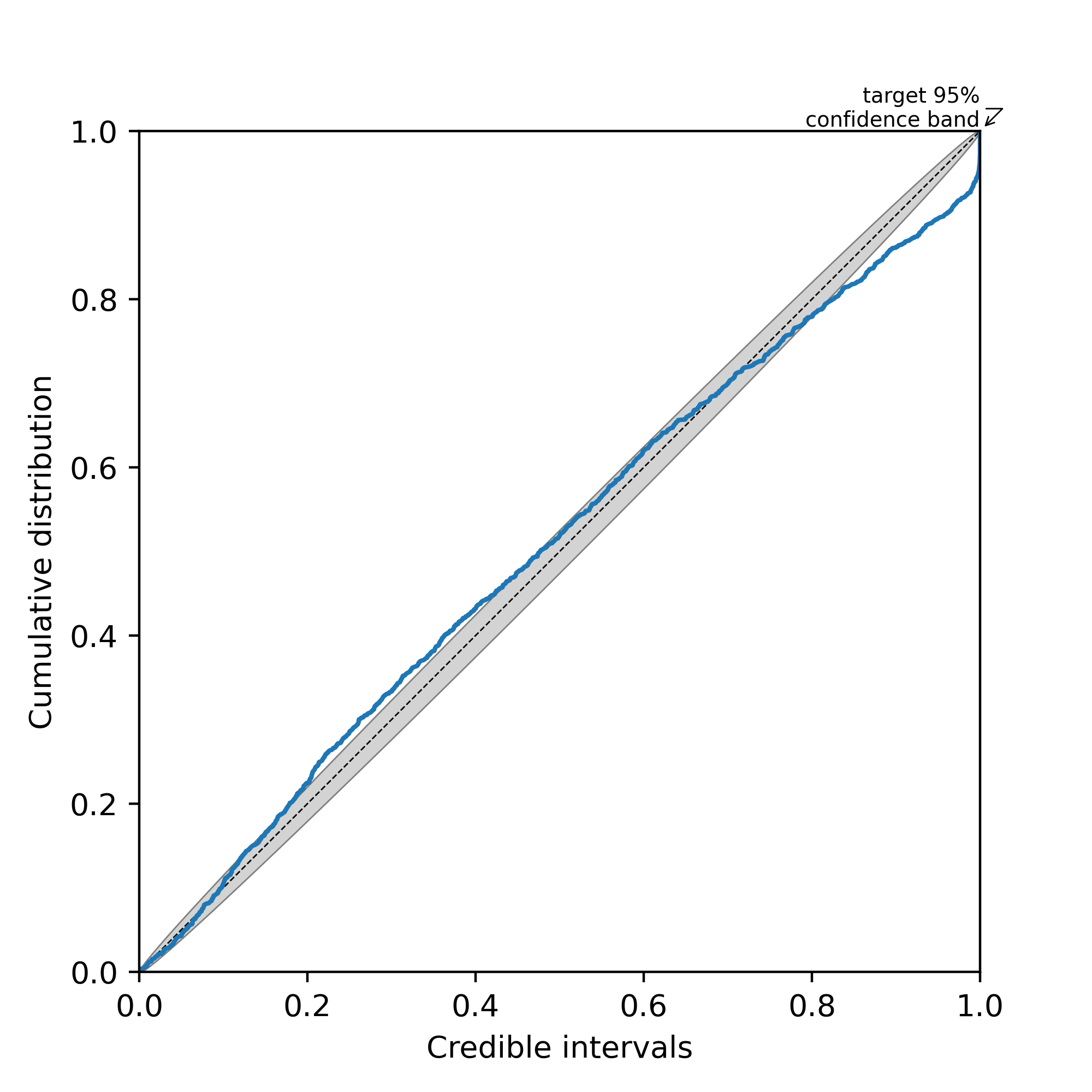}{0.32\textwidth}{(d)}
          \fig{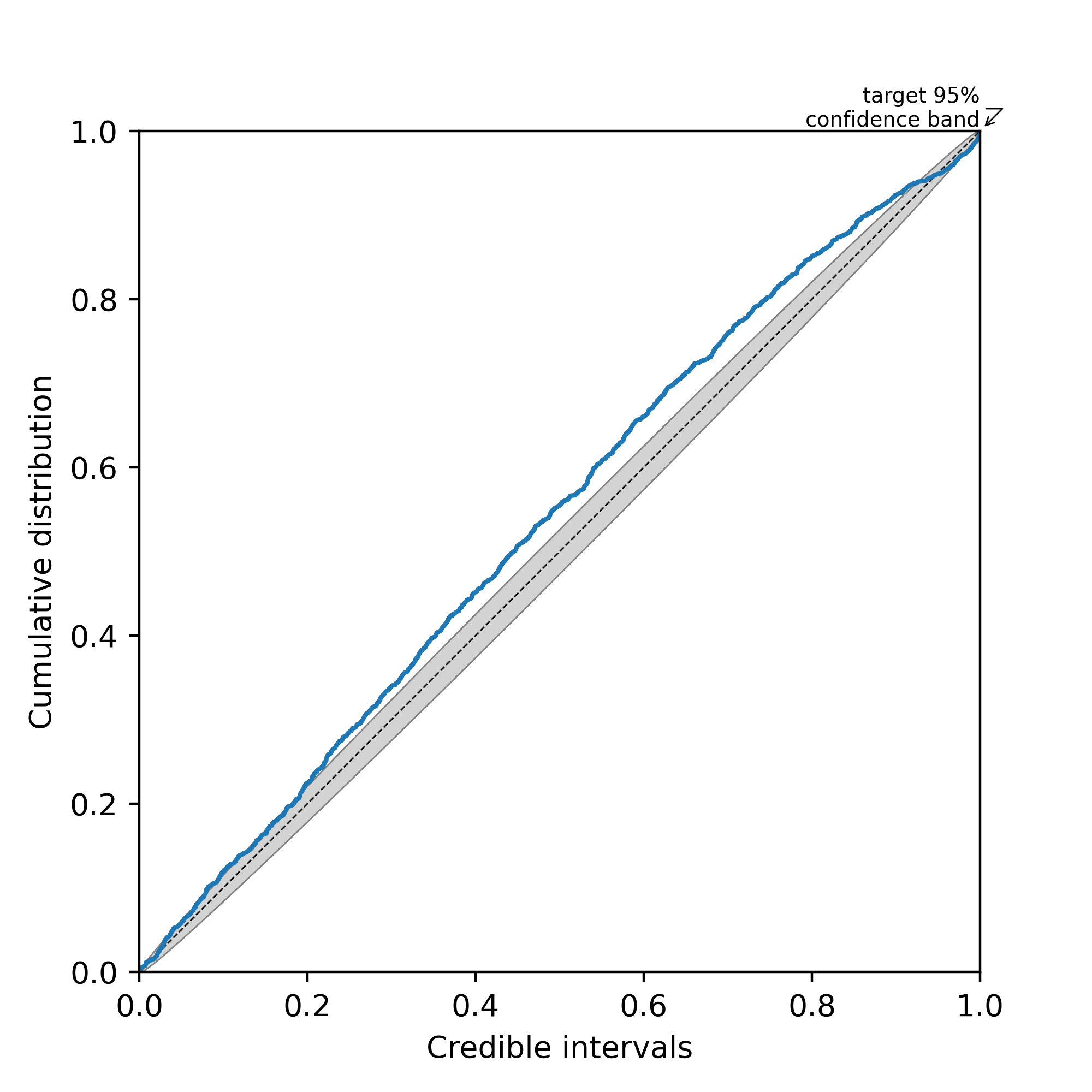}{0.32\textwidth}{(e)}
          \fig{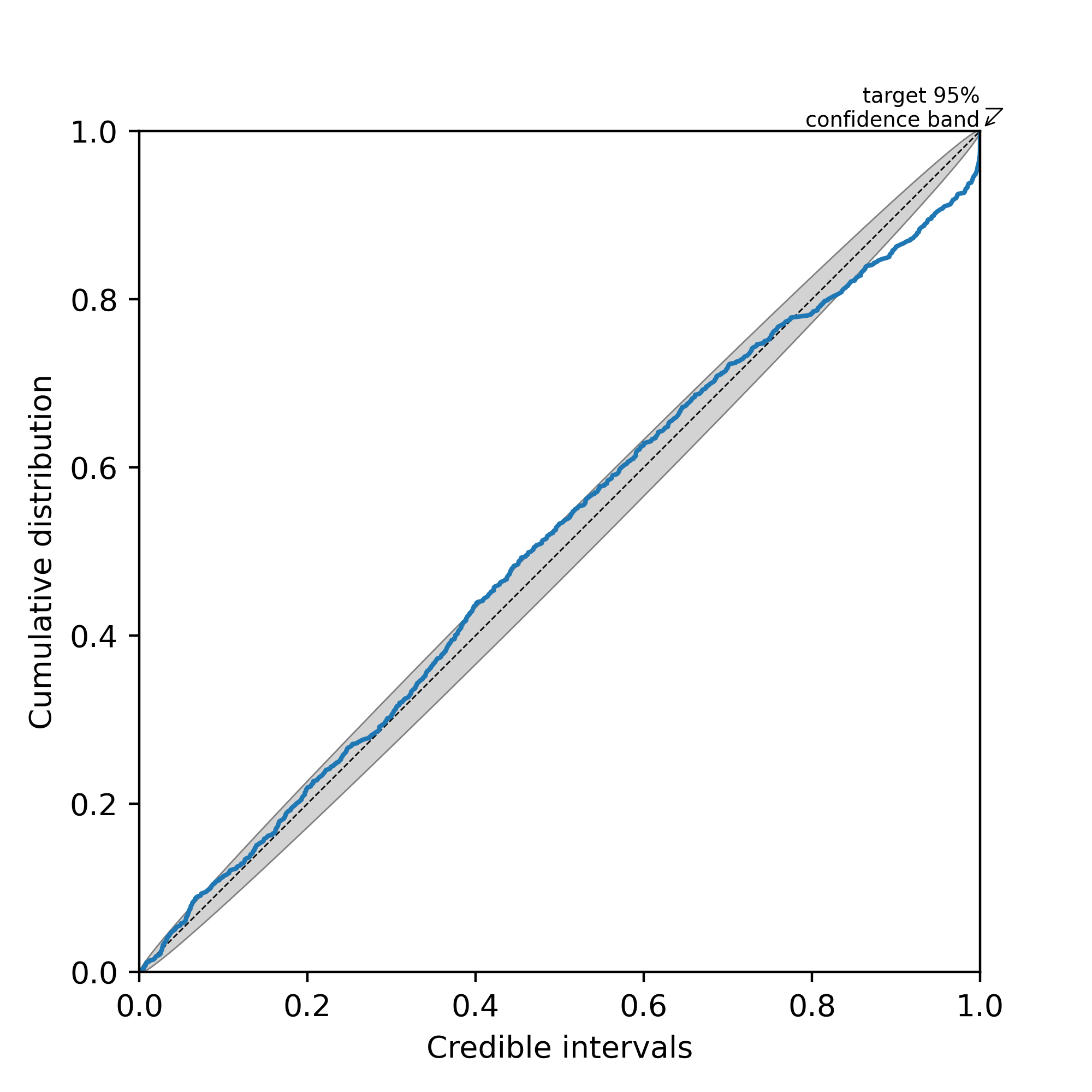}{0.32\textwidth}{(f)}}
\caption{(a) to (f): P–P plots for test samples at 0 secs, 10 secs, 15 secs, 28 secs, 44 secs and 58 secs before merger. The figure shows the cumulative distribution function of the searched probability which should lie close to the diagonal (dashed line) if \texttt{GW-SkyLocator}'s posterior distributions are accurate. The gray band shows the 95\% confidence band.}
\end{figure}

\section{Code and Data Availability}
The code for \texttt{GW-SkyLocator} can be found here: \href{https://github.com/chayanchatterjee/GW-SkyLocator}{GW-SkyLocator}. The test data for reproducing our figures can be found here: \href{https://doi.org/10.5281/zenodo.7746302}{https://doi.org/10.5281/zenodo.7746302}. The parameters for the injections used and the \texttt{BAYESTAR} results can be accessed from: \href{https://gstlal.docs.ligo.org/ewgw-data-release/index.html}{https://gstlal.docs.ligo.org/ewgw-data-release/index.html}.

\section{Discussion}
In summary, we have reported the first deep learning based approach for pre-merger sky localization of BNS sources, capable of orders of magnitude faster inference than Bayesian methods. Currently our model's accuracy is similar to \texttt{BAYESTAR} on injections with network SNR between 9 and 40 at design sensitivity. The next step in this research would be to perform similar analysis on real detector data which has non-stationary noise and glitches that may corrupt the signal and affect detection and sky localization. A possible way to improve our model's performance at high SNRs ($>$ 25) would be to use a finer angular resolution in the sky for evaluating the posteriors. We can also train different versions of the model for different luminosity distance (and hence SNR) ranges. Our long-term goal is to construct an independent machine learning pipeline for pre-merger detection and localization of GW sources. The faster inference speed of machine learning models would be crucial for electromagnetic follow-up and observation of prompt and precursor emissions from compact binary mergers. This method is also scalable and can be applied for predicting the luminosity distance of the sources pre-merger, which would help obtain volumetric localization of the source and potentially identify host galaxies of BNS mergers.

\begin{acknowledgments}
The authors would like to thank  Leo Singer, Foivos Diakogiannis, Kevin Vinsen, Amitava Datta and Manoj Kovalam for useful discussions on this work.This research was supported in part by the Australian Research Council Centre of Excellence for Gravitational Wave Discovery (OzGrav, through Project No. CE170100004). This research was undertaken with the support of computational resources from the Pople high-performance computing cluster of the Faculty of Science at the University of Western Australia. This work used the computer resources of the OzStar computer
cluster at Swinburne University of Technology. The OzSTAR program receives funding in part from the Astronomy National Collaborative Research Infrastructure Strategy (NCRIS) allocation provided by the Australian Government. This research used data obtained from the Gravitational Wave Open
Science Center (https://www.gw-openscience.org), a service of LIGO Laboratory, the LIGO Scientific Collaboration and the Virgo Collaboration. LIGO is funded by the U.S. National Science Foundation. Virgo is funded by the French Centre National de Recherche Scientifique (CNRS), the Italian Istituto Nazionale della Fisica Nucleare (INFN) and the Dutch Nikhef, with contributions by Polish and Hungarian institutes. This material is based upon work supported by NSF's LIGO Laboratory which is a major facility fully funded by the National Science Foundation. 
\end{acknowledgments}

\bibliography{sample631}{}
\bibliographystyle{aasjournal}



\end{document}